\documentclass[a4paper,12pt]{extarticle}
\usepackage{hyperref}
\usepackage[nosort]{cite}
\usepackage{comment}

\usepackage{csquotes}
\usepackage{color}
\usepackage{amssymb,amsmath,amsfonts,amsthm,fullpage,epic,eepic,float}
\usepackage[cp1251]{inputenc}
\usepackage[mathscr]{eucal}
\usepackage{amsmath}
\usepackage{amsfonts}
\usepackage{amssymb}
\usepackage{graphicx}
\usepackage{array}
\usepackage{cancel}
\usepackage{caption}
\usepackage{wrapfig}
\usepackage{secdot}
\usepackage{indentfirst}
\usepackage{mathrsfs}
\usepackage{cancel}
\usepackage{misccorr}
\usepackage{siunitx}
\usepackage{amsthm}
\usepackage[left=1cm,right=1cm,top=2cm,bottom=2cm,bindingoffset=0cm]{geometry}
\usepackage[OT1, T2A]{fontenc}
\usepackage[english]{babel}
\usepackage{rotating,indentfirst,array,varioref}
\usepackage{appendix,marginnote,tikz,pgf,mathtools}
\usepackage{setspace}
\usetikzlibrary{arrows.meta}

\usepackage{authblk}
\numberwithin{equation}{section}
\definecolor{myred}{RGB}{220,30,40}
\definecolor{myblue}{RGB}{60,70,220}
\definecolor{mygreen}{RGB}{40,170,70}
\definecolor{myyellow}{RGB}{250,180,0}
\definecolor{mydarkred}{RGB}{140,20,30}

\usepackage{tikz,pgf}
\usetikzlibrary{intersections, calc}
\def\be{\begin{equation}}
\def\ee{\end{equation}}
\def\ba{\begin{aligned}}
\def\ea{\end{aligned}}

\def\barr{\begin{array}}
\def\earr{\end{array}}

\def\<{\left(}
\def\>{\right)}

\def\l|{\left|}
\def\r|{\right|}

\def\p{\partial}

\numberwithin{figure}{section}

\counterwithin{figure}{section}
\begin{document}
\title{Monodromy and geometry of heavy-light Virasoro blocks}

\author{Mikhail Belakovskiy$^{1}$\thanks{mikhailbe@ariel.ac.il}}
\author{Vladimir Belavin$^{1}$\thanks{vladimirbe@ariel.ac.il} }

\affil{$^1$ Ariel University\\
Ramat HaGolan, 60\\ Ariel, Israel}

\maketitle
\abstract{The AdS/CFT correspondence relates gravity in anti-de Sitter space to a boundary conformal field theory, and in its AdS$_3$/CFT$_2$ instance the Virasoro symmetry of the boundary theory organizes correlation functions into conformal blocks. In the semiclassical limit these blocks are computed by lengths of geodesic networks in the bulk, most sharply in the heavy-light regime, where heavy operators source a background probed by light ones.
We relate the classical monodromy method to this bulk geometry in holographic coordinates, showing that the eigenvectors of the monodromy matrix encode the endpoints of bulk geodesics. This yields the light action and the equations determining the internal geodesic network; crucially, the internal network equations are independent of the heavy background. For two heavy operators we rederive the same equations from elementary Euclidean geometry, which provides an independent geometric check. As an application we compute the full non-vacuum 5-point HHLLL block, so far known only in the superlight approximation. More broadly, our construction gives a general framework for computing heavy-light blocks from the bulk, while at the same time fixing its threshold of computability.
}
\flushbottom 
%%%%%%%%%%%%%%%%%%%%%%%%%%%%%%%%%%%%%%%%%%%%%%%%
%%%%%%%%%%%%%%%%%%%%%%%%%%%%%%%%%%%%%%%%%%%%%%%%

\newpage

%%%%%%%%%%%%%%%%%%%%%%%%%%%%%%%%%%%%%%%%%%%%%%%%
%%%%%%%%%%%%%%%%%%%%%%%%%%%%%%%%%%%%%%%%%%%%%%%%
%%%%%%%%%%%%%%%%%%%%%%%%%%%%%%%%%%%%%%%%%%%%%%%%

\section{Introduction}
Conformal blocks are the kinematical building blocks of conformal field theory:
together with the spectrum and the OPE coefficients they assemble correlation
functions, and they are the central objects of the conformal bootstrap. In two
dimensions the symmetry is enhanced to the infinite-dimensional Virasoro algebra
\cite{BPZ, Zamolodchikov}, which fixes the blocks far more rigidly than in higher dimensions but renders
them difficult to evaluate in closed form. They become both tractable and
physically transparent in the semiclassical limit of large central charge $c$, the
regime relevant to the AdS$_3$/CFT$_2$ correspondence: there the Virasoro blocks
exponentiate \cite{Exponentiation}, $\mathfrak{F}\sim\exp\!\big(\tfrac{c}{6}f\big)$, and the
classical block $f$ acquires a dual description in three-dimensional gravity
\cite{Fitzpatrick1}--\cite{Semiclassical AdS3}.

Two complementary methods produce $f$. In the classical monodromy method one
inserts a degenerate operator, which obeys a second-order equation
$\psi''(z)+T(z)\psi(z)=0$ whose potential $T(z)$ carries the accessory parameters;
demanding trivial monodromy of $\psi$ around the relevant cycles fixes these
parameters and hence the block. In the holographic method the classical block is
given by the regularized length of a network of bulk geodesics anchored on the
boundary \cite{Fitzpatrick1}--\cite{Semiclassical AdS3} --- the worldlines of the operators propagating in AdS$_3$. The
equivalence of the two computations was established in \cite{Classical via AdS/CFT}, \cite{Monodromic vs Geodesic} and developed, via the
geodesic-network and Steiner-tree formulations, in \cite{Belavin Geiko, Classical via AdS/CFT 2, Steiner Trees, Holographic}.

Both methods are at their sharpest in the heavy-light limit. A few heavy operators, whose
dimensions scale with the central charge, backreact and source a nontrivial
background in which the remaining light operators propagate
as probe geodesics.

Working in holographic coordinates \cite{Holographic} --- the ratio of two linearly independent solutions of the
monodromy equation $\psi''(z)+T(z)\psi(z)=0$ --- we show that the eigenvectors of the monodromy matrix encode
the endpoints of the bulk geodesics. This produces at once the light action and the
algebraic equations that determine the internal geodesic network, and these
equations prove to be independent of the heavy background. The light action depends in a simple way on the solution of these algebraic equations, so solving them yields the desired block. In other words, this technique simplifies the methods developed in \cite{Holographic} and allows one to compute more complicated blocks. 

We also develop
an elementary Euclidean-geometry technique on the plane for the case of two heavy operators, which provides an independent geometric check of the
monodromy result. The resulting prescription is algorithmic as well and has two ingredients:
solving the algebraic system that fixes the endpoints of the internal network, and
reconstructing the classical action additively, as a sum of elementary vacuum
four- and five-point contributions. 

We set up the system for an arbitrary number of heavy fields with dimensions $\epsilon_h = \frac 6c \Delta_h \sim \mathcal O (1)$ and light fields with dimensions $\epsilon_l = \frac 6 c \Delta_l \sim \mathcal O (c^{-1})$, and carry
out the full procedure, action included, for the non-vacuum five-point HHLLL block,
so far available only in the superlight approximation \cite{From Global}. The same structure exposes
the threshold of analytic computability: closed-form solvability requires the network
equations to be solvable by radicals (the light sector) and the heavy background to
be available explicitly (the heavy sector), and these two independent bounds single
out the six-point HHHLLL block as the maximal analytically tractable case.

The paper is organized as follows. Section 2 develops the
monodromy approach in holographic coordinates and derives the background-independent
network equations. Section 3 presents the dual Euclidean-plane geometry and
reproduces these equations geometrically. Section 4 applies the formalism to the
four- and five-point non-vacuum blocks and discusses the general $n$-point structure
together with its computability threshold. We conclude in Section 5.

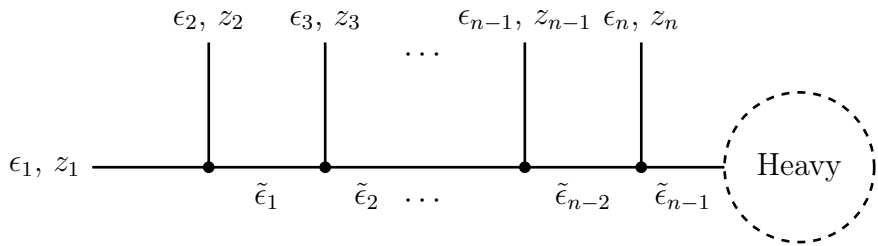
\begin{figure}[htbp]
    \centering
    \begin{tikzpicture}[scale=1.1, line width=1pt]
        % --- горизонтальная линия: epsilon_1 (нога слева) через гребёнку к Heavy ---
        \draw (0.8,0) -- (8.4,0);
        \node[left] at (0.8,0) {$\epsilon_1,\, z_1$};

        % --- Heavy blob СПРАВА (тяжёлая гребёнка, явно не рисуется) ---
        \draw[dashed] (9.3,0) circle (0.9);
        \node at (9.3,0) {Heavy};

        % --- вертикальные лёгкие ноги; epsilon_2 стоит в ПЕРВОЙ вершине (с epsilon_1) ---
        \foreach \x/\lab in {2.2/{\epsilon_2,\, z_2}, 3.6/{\epsilon_3,\, z_3},
                             6.0/{\epsilon_{n-1},\, z_{n-1}}, 7.4/{\epsilon_n,\, z_n}} {
            \fill (\x,0) circle (2pt);
            \draw (\x,0) -- (\x,1.5);
            \node[above] at (\x,1.5) {$\lab$};
        }

        % --- многоточие для произвольного числа лёгких ---
        \node at (4.8,1.35) {$\cdots$};
        \node at (4.8,-0.4) {$\cdots$};

        % --- внутренние (обменные) каналы; \tilde\epsilon_{n-1} крепится к Heavy ---
        \node[below] at (2.9,-0.05) {$\tilde\epsilon_1$};
        \node[below] at (4.1,-0.05) {$\tilde\epsilon_2$};
        \node[below] at (6.7,-0.05) {$\tilde\epsilon_{n-2}$};
        \node[below] at (7.9,-0.05) {$\tilde\epsilon_{n-1}$};
    \end{tikzpicture}
    \caption{Comb channel of the heavy-light block.}
    \label{fig:comb}
\end{figure}
%%%%%%%%%%%%%%%%%%%%%%%%%%%%%%%%%%%%%%%%%%%%%%%%%%%%%%%%%%%%%%%%%%%%%%%%%%%%%%%%%%%%%%%%%%%
%%%%%%%%%%%%%%%%%%%%%%%%%%%%%%%%%%%%%%%%%%%%%%%%%%%%%%%%%%%%%%%%%%%%%%%%%%%%%%%%%%%%%%%%%%%
%%%%%%%%%%%%%%%%%%%%%%%%%%%%%%%%%%%%%%%%%%%%%%%%%%%%%%%%%%%%%%%%%%%%%%%%%%%%%%%%%%%%%%%%%%%

\section{Monodromy}

\subsection{Standard monodromy approach}
A common, though computationally heavy, method for computing semiclassical blocks is the monodromy method \cite{Kraus Castro W_N}. The standard approach starts from the monodromy equation
\be \label{Fuchs}
\psi''(z)+T(z)\psi(z)=0,\qquad
T(z)=\sum_{i=1}^N\Big(\frac{\epsilon_i}{(z-z_i)^2}-\frac{c_i}{z-z_i}\Big),
\ee
where
$\epsilon_i=\tfrac{6}{c}\Delta_i$ and the $c_i$ are unknown accessory parameters obeying the
conformal Ward identities
\be \label{Ward}
\sum_{i=1}^N c_i=0,\qquad \sum_{i=1}^N(c_iz_i-\epsilon_i)=0,\qquad
\sum_{i=1}^N(c_iz_i^2-2\epsilon_iz_i)=0 .
\ee
These fix three combinations of the $c_i$ and leave $n-3$ of them free. The accessory parameters
are related to the classical block $f$ by
$$
c_i=-\frac {\p f}{\p z_i}.
$$
The remaining $n-3$ parameters are fixed by demanding a prescribed monodromy of the solutions of
Eq.~\ref{Fuchs} around the contour encircling the first $k$ light singularities: the corresponding
monodromy matrix $M_{12\dots k}$ must have eigenvalues
$$
\lambda^{(k-1)}_{\pm}=e^{i\pi(1\pm\sqrt{1-4\tilde \epsilon_{k-1}})}
$$
set by the intermediate dimension $\tilde\epsilon_{k-1}$ flowing in that channel.

In the heavy--light regime we split the stress tensor into a heavy background and a light part,
$T=T^{(0)}+T^{(1)}+\dots$, where $T^{(0)}$ is sourced by the heavy operators and $T^{(1)}$ is linear
in the light dimensions; the solutions of Eq.~\ref{Fuchs} are expanded likewise,
$\psi=\psi^{(0)}+\psi^{(1)}+\dots$. 

The background $T^{(0)}$ is itself a monodromy problem for the
$n_h$ heavy operators alone, with $n_h$ accessory parameters of its own. Here a sharp restriction
appears: for $n_h\le3$ the Ward identities \ref{Ward} already fix all of them --- no free
parameters remain --- so $T^{(0)}$ is known in closed form (for $n_h=2$ it is the conical-defect
background $T^{(0)}=\epsilon_h/z^2$). For $n_h\ge4$, however, $n_h-3$ accessory parameters survive
and are fixed only by the monodromy conditions, so no explicit background is
available. The heavy sector is therefore solvable in closed form only up to three heavy operators.

Given two linearly independent background solutions $\psi_{1,2}^{(0)}$, the first-order correction
is obtained by integration,
\be
\ba 
\psi_{i}^{(1)}(z)=\frac 1 W \psi_1^{(0)} \int dz \psi_2^{(0)} T^{(1)}\psi_i^{(0)}-\frac 1 W \psi_2^{(0)}\int dz \psi_1^{(0)} T^{(1)} \psi_i^{(0)}, \;\;\; 
\ea 
\ee 
where $W=\psi_1^{(0)}\psi_2^{(0)\prime}-\psi_1^{(0)\prime}\psi_2^{(0)}$ is the Wronskian. From this
we build the first-order monodromy matrix
\be 
\ba 
M_{12\dots k }^{(1)}=\frac 1 W \oint \limits_{\gamma_{12\dots k}}dz \begin{pmatrix}
    \psi_1^{(0)} T^{(1)}\psi_2^{(0)} & \psi_2^{(0)} T^{(1)}\psi_2^{(0)}\\
   - \psi_1^{(0)} T^{(1)}\psi_1^{(0)}& -\psi_2^{(0)} T^{(1)}\psi_1^{(0)}
\end{pmatrix} .
\ea 
\ee 
Imposing the eigenvalue condition on $M_{12\dots k}$ yields the equations for the accessory
parameters; solving these for the $c_i$ and integrating $c_i=-\p f/\p z_i$ reconstructs the
classical block $f$. In general, $M_{12\dots k}$ encodes the global connection data of Eq.~\ref{Fuchs},
and the resulting conditions are difficult to solve, which motivates the more direct route developed below.

\subsection{Monodromy matrix and its eigenvectors}

In the bulk each light operator is realized as a geodesic anchored on the boundary, and this geodesic meets the boundary at two points: the insertion point $z_i$ and a second point $y_i$. The monodromy $M_{\gamma_i}$ acquired on encircling the operator is precisely the bulk holonomy around this geodesic, so it is fixed by the geodesic as a whole --- by the pair of endpoints $(z_i,y_i)$, not by $z_i$ alone,
$$M_{\gamma_i}=M_{\gamma_i}(\epsilon_i,z_i,y_i).$$
(The bulk geometry behind this picture is developed in Sec.~3.) The precise way in which $M_{\gamma_i}$ records $y_i$ is what we establish below.

We will be working in the holographic coordinates developed in \cite{Holographic}. To this end we define $\mathfrak w (z) = \frac{\psi_1(z)}{\psi_2(z)}$, where $\psi_{1,2}$ are two linearly independent solutions of the monodromy problem for the heavy background, $\psi_i''+T^{(0)}\psi_i=0, \;\; i=1,2$. (Here and in what follows we omit the superscript $^{(0)}$ on the solutions $\psi_i$ for simplicity).
\\
We diagonalize the monodromy matrix:
\be \label{monodromym}
M_{\gamma_i}=\text{Id}+2\pi i \epsilon_i P_i\Lambda P_i^{-1}+\mathcal O(c^{-2}), \;\; \Lambda = \text{diag}( 1, - 1)
\ee 
Now we take a closer look at $P_i$. We claim that the columns of $P_i$ encode the ends of the geodesic as points on $\mathbb P^1$. More precisely:
$$
\begin{pmatrix}
    -1& -1\\
    \mathfrak w_i(x)& \mathfrak  w_i(y)
\end{pmatrix}
$$
To see this, we compute the monodromy matrix for the vacuum 4-point block. The block itself reads \cite{Holographic}
$$
f(z)=\epsilon_1 \ln \mathfrak w_1' +\epsilon_1 \ln\mathfrak  w_2' - 2\epsilon_1\ln(\mathfrak w_1-\mathfrak w_2)$$

$$c_1=-\epsilon_1 \frac {\mathfrak w_1''}{\mathfrak w_1'}+2\epsilon_1\frac{\mathfrak w_1'}{\mathfrak w_1-\mathfrak w_2}$$
\be 
\ba 
\frac{1}{2\pi i}\oint\limits_{\gamma_1}dz \psi_1\psi_2  T^{(1)}&=\epsilon_1(\psi_1\psi_2)'-c_1\psi_1\psi_2\\
\frac{1}{2\pi i}\oint\limits_{\gamma_1}dz \psi_1^2 T^{(1)}&=2\epsilon_1\psi_1\psi_1'-c_1\psi_1^2\\
\frac{1}{2\pi i}\oint\limits_{\gamma_1}dz \psi_2^2 T^{(1)}&=2\epsilon_1\psi_2\psi_2'-c_1\psi_2^2
\ea 
\ee
The monodromy matrix at the first order reads:
\be\label{matrix}
\frac {2\pi i \epsilon_1} W
\begin{pmatrix}
    \psi_1\psi_2'+\psi_1'\psi_2-\frac{c_1}{\epsilon_1}\psi_1\psi_2& 2\psi_2\psi_2'-\frac{c_1}{\epsilon_1}\psi_2^2\\
    -2\psi_1\psi_1'+\frac{c_1}{\epsilon_1}\psi_1^2& -\psi_1\psi_2'-\psi_2\psi_1'+\frac{c_1}{\epsilon_1}\psi_1\psi_2
\end{pmatrix}
\ee
and its eigenvectors are
$$
\begin{pmatrix}
    -1\\
   \mathfrak  w_1
\end{pmatrix}, 
\begin{pmatrix}
    -1\\
    \frac{\frac{c_1}{\epsilon_1} \psi_1-2\psi_1'}{\frac{c_1}{\epsilon_1}\psi_2-2\psi_2'}
\end{pmatrix}
$$
$$
\frac{c_1}{\epsilon_1}\psi_1-2\psi_1'=-\frac{2W}{\psi_2}+\frac{2W \mathfrak  w_1}{\psi_2(\mathfrak w_1-\mathfrak w_2)}=\frac{2W\mathfrak w_2}{\psi_2(\mathfrak w_1-\mathfrak w_2)}
$$
$$
\frac{c_1}{\epsilon_1}\psi_2-2\psi_2'=\frac{2W}{\psi_2(\mathfrak w_1-\mathfrak w_2)}
$$
Hence the eigenvectors are
$$
\begin{pmatrix}
    -1\\
   \mathfrak  w_1
\end{pmatrix}, \begin{pmatrix}
    -1\\
   \mathfrak  w_2
\end{pmatrix}
$$
Therefore, the eigenvectors of the monodromy matrix encode the ends of the geodesic. 

Consider three geodesics that merge at some point in the bulk or, in other words, the network of the 5-point vacuum block (Type-I, $\tilde\epsilon_1=\epsilon_3, \;\tilde\epsilon_2=0$). Let their ends have coordinates $z_1, y_1$, $z_2, y_2$, $z_3, y_3$. We choose a base point and consider the three fundamental cycles that encircle $z_1, z_2, z_3$. Generally speaking, the cycle $\gamma_1\circ\gamma_2\circ\gamma_3$ is not contractible on the boundary due to the presence of the heavy fields, but it is contractible in the bulk. The bulk metric generated by $T$ provides a flat connection, because it solves the Einstein equations. We can therefore pull the cycle into the bulk while preserving the initial holonomy, which turns out to be trivial. This gives the following equation:

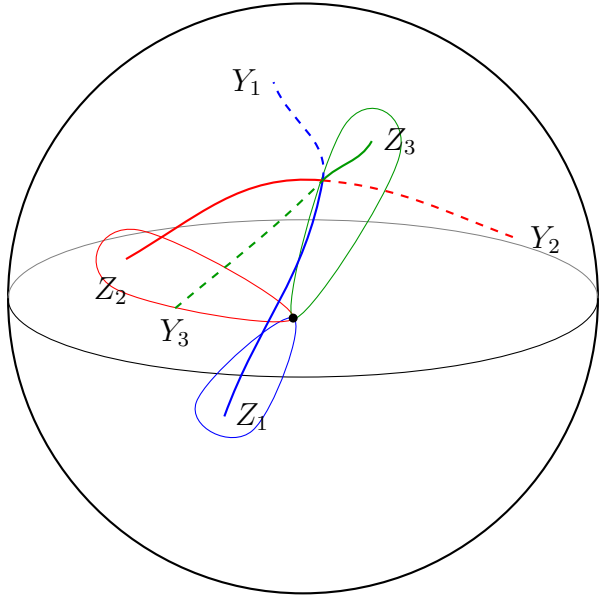
\begin{figure}[htbp]
    \centering
    \begin{tikzpicture}[scale=1.3]
        % 1. Граница AdS3 (Сфера)
        \draw[thick] (0,0) circle (3);

        % 2. Экватор
        \draw[very thin, gray] (3,0) arc (0:180:3 and 0.8);
        \draw[thin] (3,0) arc (360:180:3 and 0.8);

        % 3. Координаты
        \coordinate (Center) at (0.2, 1.2);
        \coordinate (Z1)  at (-0.8, -1.2);
        \coordinate (Z2)  at (-1.8, 0.4);
        \coordinate (Y1)  at (-0.3, 2.2);
        \coordinate (Y2)  at (2.2, 0.6);
        \coordinate (Z11) at (0.7, 1.6);
        \coordinate (Z12) at (-1.3, -0.1);
        \coordinate (Inf) at (0.5, 2.5);
        \coordinate (Zero) at (0.0, -2.0);
        \coordinate (BP)  at (-0.1, -0.2);   % базисная точка

        % 4. Три контура из ОДНОЙ базисной точки BP (все проходят через BP)
        \draw[blue, thin] plot[smooth cycle, tension=0.8] coordinates {
            (BP) (-1.085,-1.055)  (-0.515,-1.345) };
        \draw[red, thin] plot[smooth cycle, tension=0.8] coordinates {
            (BP) (-1.92,0.104) (-1.68,0.697) };
        \draw[green!60!black, thin] plot[smooth cycle, tension=0.8] coordinates {
            (BP) (0.98,1.445) (0.42,1.755) };
        \fill (BP) circle (1.3pt);

        % 5. Геодезики (ОРИГИНАЛ, не тронуты)
        \draw[blue, thick] (Z1) to[out=70,in=260] (Center);
        \draw[blue, thick, dashed] (Center) to[out=80,in=-70] (Y1);
        \draw[red, thick] (Z2) to[out=30,in=175] (Center);
        \draw[red, thick, dashed] (Center) to[out=-5,in=160] (Y2);
        \draw[green!60!black, thick, dashed] (Z12) to[out=40,in=225] (Center);
        \draw[green!60!black, thick] (Center) to[out=45,in=240] (Z11);

        % 6. Подписи
        \node[right] at (Z1) {$Z_1$};
        \node[below, xshift=-2mm, yshift=-1mm] at (Z2) {$Z_2$};
        \node[left]  at (Y1) {$Y_1$};
        \node[right] at (Y2) {$Y_2$};
        \node[below] at (Z12) {$Y_3$};
        \node[right] at (Z11) {$Z_3$};
    \end{tikzpicture}
    \caption{Three geodesics and three contours}
    \label{riemannsphere}
\end{figure}

\be 
M_{\gamma_{z_1}}M_{\gamma_{z_2}}M_{\gamma_{z_3}}=M_{123}=\text{Id}
\ee 
and at the first order
\be \label{Id}
\sum\limits_{i=1}^3\epsilon_iP_i\Lambda P_i^{-1}=0
\ee

In terms of the holographic coordinate $\mathfrak w=\frac{\psi_1^{(0)}(z)}{\psi_2^{(0)}(z)}$, 
Eq.~\ref{Id} yields exactly three independent equations:
\be \label{3equations}
\ba 
\sum\limits_{i=1}^3 \epsilon_i \frac{\mathfrak w(z_i)+\mathfrak w(y_i)}{\mathfrak w(z_i)-\mathfrak w(y_i)}=0\\
\sum\limits_{i=1}^3 \epsilon_i \frac{1}{\mathfrak w(z_i)-\mathfrak w(y_i)}=0\\
\sum\limits_{i=1}^3 \epsilon_i \frac{\mathfrak w(z_i)\mathfrak w(y_i)}{\mathfrak w(z_i)-\mathfrak w(y_i)}=0
\ea 
\ee 
These equations can be solved explicitly,
\be \label{solutions}
\ba
\mathfrak w(y_i)=-\prod\limits_{j=1}^3\mathfrak {w}_j \frac{2\epsilon_i \mathfrak w_i^{-1}-(\epsilon_i-\epsilon_{i+1}+\epsilon_{i+2})\mathfrak w_{i+1}^{-1}-(\epsilon_i+\epsilon_{i+1}-\epsilon_{i+2})\mathfrak w_{i+2}^{-1}}{2\epsilon_i \mathfrak w_i-(\epsilon_i-\epsilon_{i+1}+\epsilon_{i+2})\mathfrak w_{i+1}-(\epsilon_i+\epsilon_{i+1}-\epsilon_{i+2})\mathfrak w_{i+2}}, \;\; \mathfrak w_i=\mathfrak w(z_i) 
\ea 
\ee 

and brought to the following invariant form:
\be\ba \label{cross}
(\mathfrak w(z_i), \mathfrak w(z_{i+1}); \mathfrak w({y}_{i+2}), \mathfrak w(z_{i+2}))=-\frac{-\epsilon_i+\epsilon_{i+1}+{\epsilon}_{i+2}}{\epsilon_i-\epsilon_{i+1}+{\epsilon}_{i+2}}, \\
z_{i+3}=z_i,\; y_{i+3}=y_i, \;\epsilon_{i+3}=\epsilon_i
\ea\ee 
First, note that \ref{cross} does not depend on the fundamental matrix, as it should not. Second, note that the argument applies to any cubic vertex of a geodesic network, not only to external insertions. Since the bulk connection is flat, the holonomy around any geodesic of the network is well defined and depends only on the homotopy class of the contour, with eigenvalues set by the dimension carried by that line --- $\epsilon_i$ for external and $\tilde \epsilon_i$ for internal lines. The contractibility argument above therefore holds at every cubic vertex, so the relations \ref{cross} apply vertex-wise to the entire network for an arbitrary heavy background. 

We now turn to the action for the 5-point vacuum block. Solving \ref{3equations}, we substitute the solutions \ref{solutions} into \ref{monodromym} and obtain the equation for the accessory parameter:
\be 
\ba 
\frac 1 W (\epsilon_1(\psi_1\psi_2)'-c_1\psi_1\psi_2)=(\epsilon_1-\epsilon_2+\epsilon_3)\frac{\mathfrak w_1}{\mathfrak w_3-\mathfrak w_1}+\frac{\epsilon_2 \mathfrak w_1-\epsilon_3\mathfrak w_1+\epsilon_1 \mathfrak w_2}{\mathfrak w_2-\mathfrak w_1}\Rightarrow \\
c_1=\epsilon_1\frac{(\psi_1\psi_2)'}{\psi_1\psi_2}-\frac{W}{\psi_1\psi_2}\Big((\epsilon_1-\epsilon_2+\epsilon_3)\frac{\mathfrak w_1}{\mathfrak w_3-\mathfrak w_1}+\frac{\epsilon_2\mathfrak w_1-\epsilon_3\mathfrak w_1+\epsilon_1 \mathfrak w_2}{\mathfrak w_2-\mathfrak w_1}\Big)
\ea
\ee
But $\psi_1\psi_2=\frac{\mathfrak wW}{\mathfrak w'}$, and therefore
\be 
\ba 
f=\epsilon_1\ln \frac{\mathfrak w'_1}{\mathfrak w_1}+\int \frac{d\mathfrak w_1}{\mathfrak w_1}\Big((\epsilon_1-\epsilon_2+\epsilon_3)\frac{\mathfrak w_1}{\mathfrak w_3-\mathfrak w_1}+\frac{\epsilon_2\mathfrak w_1-\epsilon_3\mathfrak w_1+\epsilon_1 \mathfrak w_2}{\mathfrak w_2-\mathfrak w_1}\Big)=\\
=\epsilon_1 \ln \mathfrak w_1'-(\epsilon_1+\epsilon_2-\epsilon_3)\ln(\mathfrak w_1-\mathfrak w_2)-(\epsilon_1-\epsilon_2+\epsilon_3)\ln(\mathfrak w_1-\mathfrak w_3)+C(\mathfrak w_2, \mathfrak w_3)
\ea 
\ee 
Finally,
\be 
f(z)=\bigg(\sum\limits_{i=1}^3\epsilon_i\ln \mathfrak w_i'-(\epsilon_1+\epsilon_2-\epsilon_3)\ln(\mathfrak w_1-\mathfrak w_2)-(\epsilon_1-\epsilon_2+\epsilon_3)\ln(\mathfrak w_1-\mathfrak w_3)-(-\epsilon_1+\epsilon_2+\epsilon_3)\ln(\mathfrak w_2-\mathfrak w_3)\bigg)\bigg|_{\mathfrak w_j = \mathfrak w (z_j)}
\ee
Therefore
\be
f(\mathfrak w)=-(\epsilon_1+\epsilon_2-\epsilon_3)\ln(\mathfrak w_1-\mathfrak w_2)-(\epsilon_1-\epsilon_2+\epsilon_3)\ln(\mathfrak w_1-\mathfrak w_3)-(-\epsilon_1+\epsilon_2+\epsilon_3)\ln(\mathfrak w_2-\mathfrak w_3)
\ee
and the corresponding action in the bulk is
\be \label{action monodromy}
f(\theta)=\bigg(\sum\limits_{i=1}^3\epsilon_i\ln\mathfrak w_i-(\epsilon_1+\epsilon_2-\epsilon_3)\ln(\mathfrak w_1-\mathfrak w_2)-(\epsilon_1-\epsilon_2+\epsilon_3)\ln(\mathfrak w_1-\mathfrak w_3)-(-\epsilon_1+\epsilon_2+\epsilon_3)\ln(\mathfrak w_2-\mathfrak w_3)\bigg)\bigg|_{\mathfrak w_j=e^{i\theta_j}}
\ee
This action is additive in the bulk. 

%%%%%%%%%%%%%%%%%%%%%%%%%%%%%%%%%%%%%%%%%%%%%%%%%%%%%%%%%%%%%%%%%%%%%%%%%%%%%%%%%%%%%%%%%%%%%%%%%%%%%%%%%%%%%%%%%%%%%%%%%%%%%%%%%%%%%%%%%%%%%%%%%%%%%%%%%%%%%%%%%%%%%
%%%%%%%%%%%%%%%%%%%%%%%%%%%%%%%%%%%%%%%%%%%%%%%%%%%%%%%%%%%%%%%%%%%%%%%%%%%%%%%%%%%%%%%%%%%%%%%%%%%%%%%%%%%%%%%%%%%%%%%%%%%%%%%%%%%%%%%%%%%%%%%%%%%%%%%%%%%%%%%%%%%%%
%%%%%%%%%%%%%%%%%%%%%%%%%%%%%%%%%%%%%%%%%%%%%%%%%%%%%%%%%%%%%%%%%%%%%%%%%%%%%%%%%%%%%%%%%%%%%%%%%%%%%%%%%%%%%%%%%%%%%%%%%%%%%%%%%%%%%%%%%%%%%%%%%%%%%%%%%%%%%%%%%%%%%
\section{Geometry}
In this section we develop a geometric technique for computing heavy-light blocks with two equal heavy fields. This technique provides an independent derivation of ~\ref{cross} and \ref{action monodromy}. 
\subsection{Bulk description}

The heavy part of the stress tensor is given by $T^{(0)}=\frac{\epsilon_h}{z^2}$. 
The bulk treatment was described in detail in \cite{Semiclassical, Classical via AdS/CFT, Semiclassical AdS3, Belavin Geiko},  so we only briefly recall the computation. The background metric in AdS$_3$ is a solution of the Einstein equations in the bulk with given boundary conditions \cite{Semiclassical}, and it is given by 
$$
ds^2=\frac{\alpha^2}{\cos^2\rho}(-dt^2+\sin^2\rho d\phi^2+\frac{1}{\alpha^2}d\rho^2),
$$
where the parameter $\alpha$ reflects the presence of the two heavy fields:  $\alpha =\sqrt{1-\frac{24 \Delta_h}{c}}$. We assume $\alpha^2>0$.
The length of a curve, with the corresponding parameterization, reads
$$
L=\int d\tau \sqrt{g_{\mu\nu}\dot x_\mu\dot x_\nu}, \;\;\;g_{\mu\nu}\dot x_\mu\dot x_\nu=1
$$
The conjugate momenta are defined as $p_i=\frac{\partial L}{\partial \dot x_i}$, and since $t, \phi$ are cyclic coordinates, $p_t=\text {const}, p_\phi = \text{const}$ on-shell.
 We will use redefined coordinates and momenta: $\alpha\phi\to \theta,\;\;\frac {p_\phi}\alpha \to s$, \;\;$z^\alpha \to  \mathfrak w$.
 The parameterization condition can be written explicitly in terms of the momenta 
 \be \label{parameterization}
 p_\rho^2\cos^2\rho+s^2\cot^2\rho=1
\ee 
so we can introduce an imaginary exponential: 
$$
e^{i\mu} = p_\rho\cos\rho+i s \cot\rho, \;\; \mu\in [0, 2\pi)
$$
We now want to find the action and momentum associated with a geodesic that starts on the boundary and reaches a point in the bulk. Using the explicit expression $p_\rho=\frac{\dot\rho}{\cos^2\rho}$, we can rewrite \ref{parameterization} as
\be 
d\lambda=\pm \frac{d\rho}{\cos\rho\sqrt{1-s^2\cot^2\rho}},
\ee
which integrates to 
\be\label{action0}
L=\int\limits_{\lambda_1}^{\lambda_2} d \lambda=\log\frac{x}{\sqrt{1+x^2}+\sqrt{1-s^2x^2}}\Big|_{x_1}^{x_2}, \;\; x=\cot\rho
\ee
To obtain the geodesic equation, we express
$\dot{\theta}=s\cot^2\rho$,
which gives 
$$
d\theta=\pm\frac{\cos\rho}{\sin^2\rho\sqrt{1-s^2\cot^2\rho}},
$$
and after integration we obtain the geodesic equation
\be \label{geodesic}
e^{i\Delta\theta}=\Big(\sqrt{1-s^2x^2}-is \sqrt{1+x^2}\Big)\Big|_{x_1}^{x_2}
\ee

%%%%%%%%%%%%%%%%%%%%%%%%%%%%%%%%%%%%%%%%%%%%%%%%%%%%%%%%%%%%%%%%%%%%%%%%%%%%%%%%%%%%%%%%%%%%%%%%%%%%%%%
%%%%%%%%%%%%%%%%%%%%%%%%%%%%%%%%%%%%%%%%%%%%%%%%%%%%%%%%%%%%%%%%%%%%%%%%%%%%%%%%%%%%%%%%%%%%%%%%%%%%%%%
%%%%%%%%%%%%%%%%%%%%%%%%%%%%%%%%%%%%%%%%%%%%%%%%%%%%%%%%%%%%%%%%%%%%%%%%%%%%%%%%%%%%%%%%%%%%%%%%%%%%%%%
%%%%%%%%%%%%%%%%%%%%%%%%%%%%%%%%%%%%%%%%%%%%%%%%%%%%%%%%%%
%%%%%%%%%%%%%%%%%%%%%%%%%%%%%%%%%%%%%%%%%%%%%%%%%%%%%

\subsection{Methodology}
We consider now a geodesic with one endpoint on the boundary at the coordinate $\mathfrak w$. From \ref{geodesic} we can extract the expression for the angular momentum
\be \label{momentum}
s=\Big|\frac{(\mathfrak w-e^{i \theta})(\mathfrak w+e^{i\theta})}{(\mathfrak w-e^{i\theta}\tan\frac\rho 2)(\mathfrak w-e^{i\theta}\cot\frac\rho 2)}\Big|
\ee
and rewrite the regularized geodesic length as 
\be \label{length}
L=\ln\frac{\sin\text{Arg} {\frac{e^{i\theta}}{\mathfrak w}}}{s\cot\rho}
\ee
We now want to build a geometric interpretation of these formulae. Observe that $s$ is built from the product and ratio of four segments in the complex plane. To this end, we introduce the following points and notation:

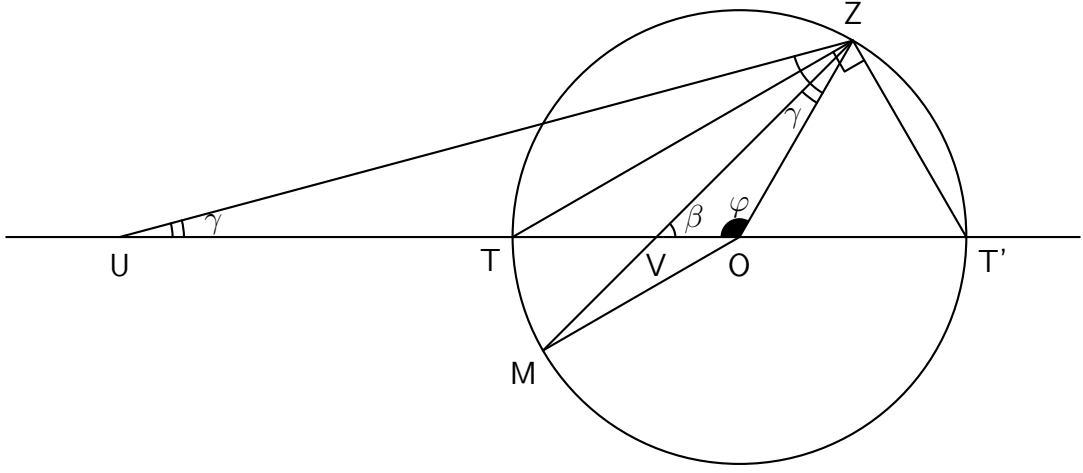
\begin{figure}[htbp] 
\centering
\begin{tikzpicture}[thick, line cap=round, line join=round]

    % --- Геометрические параметры ---
    \def\R{3} % Радиус окружности

    % --- Координаты основных точек ---
    \coordinate (O) at (0,0);
    \coordinate (Z) at (60:\R);
    \coordinate (L) at (-\R,0); % Левое пересечение с окружностью
    \coordinate (R) at (\R,0);  % Правое пересечение с окружностью
    
    % V - пересечение биссектрисы угла OZL с осью X
    \coordinate (V) at ({1.5 - 1.5*sqrt(3)}, 0);
    
    % M - точка на окружности (продолжение отрезка ZV)
    \coordinate (M) at (210:\R);
    
    % U - точка на оси X такая, что угол UZL равен гамма (15 градусов)
    \coordinate (U) at ({-3 - 3*sqrt(3)}, 0);

    % --- Отрисовка линий ---
    % Главная горизонтальная ось
    \draw ($(U) + (-1.5,0)$) -- ($(R) + (1.5,0)$);

    % Окружность
    \draw (O) circle (\R);

    % Соединительные отрезки
    \draw (U) -- (Z);
    \draw (L) -- (Z);
    \draw (M) -- (Z); % Линия MZ проходит ровно через V
    \draw (O) -- (Z);
    \draw (R) -- (Z);

    % --- Углы и дуги ---

    % Угол \alpha у точки O (залитый черным сектор)
    \fill (O) -- (180:0.25) arc (180:60:0.25) -- cycle;
    \node at (90:0.4) {$\varphi$}; % Отодвинуто на пустое пространство

    % Угол \beta Ау точки V
    \draw (V) ++(0:0.25) arc (0:45:0.25);
    \node at ($(V) + (25:0.55)$) {$\beta$}; % Отодвинуто на пустое пространство

    % Угол \gamma у точки U (двойная дуга)
    \draw (U) ++(0:0.7) arc (0:15:0.7);
    \draw (U) ++(0:0.85) arc (0:15:0.85);
    \node at ($(U) + (7.5:1.25)$) {$\gamma$}; % Вынесено дальше, чтобы не задевать узкий угол

    % Углы у точки Z
    % Непрерывная внутренняя дуга для трех смежных углов (от 195° до 240°)
    \draw (Z) ++(195:0.8) arc (195:240:0.8);
    
    % Внешняя дуга для угла \gamma у точки Z (от 195° до 210°)
    \draw (Z) ++(225:0.95) arc (225:240:0.95);
    \node at ($(Z) + (232.5:1.3)$) {$\gamma$}; % Отодвинуто дальше от пучка линий
    
    % Прямой угол у точки Z (между ZL и ZR)
    \draw ($(Z) + (210:0.3)$) -- ++(300:0.3) -- ++(30:0.3);

    % --- Подписи точек ---
    \node[below, yshift=-2pt] at (U) {\textsf{U}};
    \node[below, yshift=-2pt] at (V) {\textsf{V}};
    \node[below, yshift=-2pt] at (O) {\textsf{O}};
    \node[above, yshift=2pt] at (Z) {\textsf{Z}};
    \node[below left, xshift=2pt] at (M) {\textsf{M}};
    
    % Добавленная подпись точки L (слева сверху от пересечения, чтобы не задевать ось)
    \node[below left] at (L) {\textsf{T}};
    \node[below right] at (R) {\textsf{T'}};
    \draw (O)--(M);
\end{tikzpicture}
\caption{A plane configuration}
\label{ZOV}
\end{figure}

$Z=\mathfrak w, T=e^{i\theta}, T'=-e^{i\theta}, U=e^{i\theta}\cot\frac \rho 2, V=e^{i\theta}\tan\frac \rho 2$. The points $Z, T, T'$ lie on the unit circle $\omega$. We define $M\in (ZV)\cap\omega$ and the angles $\angle ZOT=\varphi, \angle ZVO = \beta, \angle OZV=\gamma$. Then $s\cot\rho=\frac{ZT\cdot ZT'}{2 ZU\cdot ZV}(OU^2-OV^2)$. First, we notice that $\triangle ZOV\sim\triangle UOZ$, so that $\angle UZO=\beta, \angle ZUO=\gamma$. 

Therefore, $\frac{ZU}{\sin\varphi}=\frac{ZV}{\sin\gamma}$ and $\frac{ZV}{\sin\varphi}=\frac{OZ}{\sin\beta}=\frac{1}{\sin\beta}$ and $\frac{OU}{\sin\beta}=\frac{OZ}{\sin\gamma}=\frac{1}{\sin\gamma}$, $OU=\frac{1}{OV}=\frac{\sin\beta}{\sin\gamma}$. We then get 
$$
s\cot\rho=\frac{(2\sin\frac\varphi 2)(2\cos\frac\varphi2)\sin\beta\sin\gamma}{\sin^2\varphi}\frac{\sin^2\beta-\sin^2\gamma}{2\sin\beta\sin\gamma}
$$
Since $\varphi+\beta+\gamma =\pi$, this becomes 
$$
s\cot\rho=\frac{\sin^2\beta-\sin^2\gamma}{\sin(\beta+\gamma)}=\sin(\beta-\gamma)=\sin\angle TOM.
$$
Now $\mu=\pm\angle TOM$ or $\mu=\pi\mp\angle TOM$. But it is clear that $e^{i\theta}e^{i\mu}$ must lie within the angle $\angle TZO$, so 
\be \label{mu}
M=e^{i\theta}e^{i\mu}
\ee
It is now easy to obtain the expression for the regularized length: 
\be \label{action}
L=\ln \frac{\sin\varphi}{\sin(\beta-\gamma)}=\ln\frac{\sin\angle ZOV}{\sin\angle TOM}=\ln\frac{ZV}{MV}
\ee
In what follows we use \ref{mu} and \ref{action} to compute heavy-light conformal blocks. 
%%%%%%%%%%%%%%%%%%%%%%%%%%%%%%%%%%%%%%%%%%%%%%%%%%%%%%%%%%%%%%%%%%%%%%%%%%%%%%%%%%%%%%%%%%%%%%%%%%%%%%%%%%%%%%%%%%%%%%%%%%%%%%%%%%%%%%%%%%%%%%%%%%%%%%%%%%%%%%%%%%%%%%%%%%%%%%%%%%%%%%%%%%%%%%%%%%%%%%%%%%%%%%%%%%%%%%%%%%%%%%%%%%%%%%%%%%%%%%%%%%%%%%%%%%%%%%%%%%%%%%%%%%%%%%%%%%%%%%%%%%%%%%%%%%%%%%%%%%%%%%%%%%%%%%%%%%%%%%%%%%%%%%%%%%%%%%%%%%%%%%%%%%%%%%%%%%%%%%%%%%%%%%%%%%%%%%%%%%%%%%%%%%%%%%%%%%%%%%%%%%%%%%%%
\subsection{4-point identity block}
Here we demonstrate how the proposed method works for computing a simple heavy-light vacuum four-point block. We pick an arbitrary point on the geodesic that connects the points $z_1$ and $z_2$, and consider the mapping that sends this point $(\rho,\theta)$ to $V=V(\rho, \theta)$. We find the two intersections $M_{1,2}\in \omega\cap (Z_{1,2}V)$. Notice that the point $(\rho, \theta)$ can be treated as a point where two different geodesics merge; this leads to the conclusion that the corresponding numbers satisfy $e^{i\mu_1}+e^{i\mu_2}=0$, which means that the points $M_1$ and $M_2$ are diametrically opposite. Hence $|M_1M_2|=2$, independently of the positions of the operators $Z_1, Z_2$. 
\begin{figure}[htbp]
\centering
\begin{tikzpicture}[scale=0.7]

    % ==========================================
    % Левый круг (AdS)
    % ==========================================
    % Заголовок
    \node at (0, 3.5) {\textsf{AdS}};
    
    % Граница диска
    \draw[thick] (0,0) circle (3);
    
    % Координаты точек на границе
    \coordinate (z1L) at (225:3);
    \coordinate (z2L) at (330:3);
    
    % Подписи точек
    \node[below left] at (z1L) {$Z_1$};
    \node[below right] at (z2L) {$Z_2$};
    
    % Кривая (геодезическая) между z1 и z2
    % Используем bend left для имитации дуги окружности, ортогональной границе
    \draw[thick] (z1L) to[bend left=50] coordinate[pos=0.25] (rho) (z2L);
    
    % Точка (rho, theta) на кривой
    \fill (rho) circle (1.5pt);
    \node[below right] at (rho) {$(\rho, \theta)$};

    % ==========================================
    % Центральная стрелка (стиль \mapsto)
    % ==========================================
    % Рисуем тонкую стрелку |-> от x=3.8 до x=5.2
\draw[->, >=Latex] (3.8,0) -- (5.0,0);

    % ==========================================
    % Правый круг (Комплексная плоскость C)
    % ==========================================
    \begin{scope}[shift={(9,0)}]; % Сдвигаем правый круг на 9 единиц вправо
        
        % Заголовок
        \node at (0, 3.5) { $\mathbb{C}$};
        
        % Граница диска
        \draw[thick] (0,0) circle (3);
        
        % Начало координат O
        \coordinate (O) at (0,0);
        \fill (O) circle (1.5pt);
        \node[above, yshift=2pt] at (O) {$O$};
        
        % Координаты точек на границе
        \coordinate (M1) at (15:3);
        \coordinate (M2) at (195:3);
        \coordinate (z1R) at (225:3);
        \coordinate (z2R) at (350:3);
        
        % Подписи точек на границе
        \node[right] at (M1) {$M_1$};
        \node[left] at (M2) {$M_2$};
        \node[below left] at (z1R) {$Z_1$};
        \node[right] at (z2R) {$Z_2$};
        
        % Пунктирная линия M1 - M2
        \draw[dashed] (M2) -- (M1);
        
        % Сплошные хорды (z1 -> M1 и z2 -> M2)
        % Задаем им имена для вычисления точки пересечения
        \draw[thick, name path=L1] (z1R) -- (M1);
        \draw[thick, name path=L2] (z2R) -- (M2);
        
        % Вычисление точки пересечения V
        \path[name intersections={of=L1 and L2, by=V}];
        \fill (V) circle (2.5pt);
        \node[below, yshift=-2pt] at (V) {$V$};
        
    \end{scope}
\end{tikzpicture}
\caption{AdS$\to$ $\mathbb C$ mapping}
 \label{4vac}
\end{figure}

Now we write down the similarity of the triangles:
$\triangle Z_1VZ_2\sim \triangle M_2VM_1\Rightarrow{}$ $\frac{Z_1V}{VM_1}\frac{Z_2V}{VM_2}=\frac{Z_1V}{VM_2}\frac{Z_2V}{VM_1}=\Big(\frac{Z_1Z_2}{M_1M_2}\Big)^2\sim \frac{(\mathfrak w_1-\mathfrak w_2)^2}{\mathfrak w_1 \mathfrak w_2}$, where we omitted an inessential factor that does not depend on the coordinates. 
The action then reads:
$$
f(\theta)=-S=\epsilon_1\ln\mathfrak w_1 +\epsilon_1\ln\mathfrak w_2-2\epsilon_1 \ln (\mathfrak w_1-\mathfrak w_2)=\epsilon_1\ln\frac{z_1^\alpha z_2^\alpha}{(z_1^\alpha-z_2^\alpha)^2}
$$
and the conformal block itself is:
$$
\mathfrak F=e^{\frac c6 f(z)} \sim \frac{z_1^{\Delta_1(\alpha-1)} z_2^{\Delta_1(\alpha-1)}}{(z_1^\alpha-z_2^\alpha)^{2\Delta_1}}
$$
\subsection {Approach to an inner vertex}
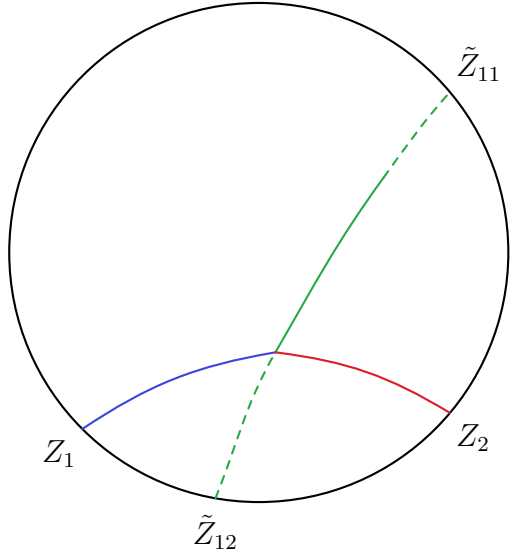
\begin{figure}[htbp] 
\centering
\begin{tikzpicture}[scale=1.1, line cap=round, line join=round, thick]
    % Базовый радиус уменьшен (R=2.2 вместо 4), 
    % чтобы размер шрифта казался в ~1.8 раза больше относительно круга.
    \def\R{3}

    % Граница круга (AdS)
    \draw[thick] (0,0) circle (\R);

    % === Координаты внутренних точек ===
    \coordinate (V1) at (0.2, -1.2); % Центральная вершина слияния
    \coordinate (V2) at (1.5, 0.9);  % Точка перехода зеленой линии (сплошная -> пунктир)

    % === Координаты точек на границе ===
    \coordinate (Z1)  at (225:\R);
    \coordinate (Z12) at (260:\R);
    \coordinate (Z2)  at (320:\R);
    \coordinate (Z11) at (40:\R);

    % === Отрисовка линий ===

    % 1. Зеленая линия (разбита на 3 сегмента для смены стилей)
    % Углы касательных строго монотонно убывают: 68° -> 60° -> 54° -> 48°,
    % что гарантирует идеальную гладкость (без изломов на стыках) и легкую выпуклость к центру.
    \draw[mygreen, dashed] (Z12) to[out=68, in=240] (V1);
    \draw[mygreen]         (V1)  to[out=60, in=234] (V2);
    \draw[mygreen, dashed] (V2)  to[out=54, in=228] (Z11);

    % 2. Синяя дуга (от Z1 к V1)
    % bend left аккуратно выгибает линию к центру (0,0)
    \draw[myblue] (Z1) to[bend left=12] (V1);

    % 3. Красная дуга (от V1 к Z2)
    % bend left аккуратно выгибает линию к центру (0,0)
    \draw[myred] (V1) to[bend left=12] (Z2);

    % === Подписи точек ===
    % Используем стандартный математический шрифт с индексами для академического вида
    \node[below left, xshift=2pt]   at (Z1)  {$Z_1$};
    \node[below, yshift=-2pt]       at (Z12) {$\tilde Z_{12}$};
    \node[below right, xshift=-2pt] at (Z2)  {$Z_2$};
    \node[above right, xshift=-2pt] at (Z11) {$\tilde Z_{11}$};

\end{tikzpicture}
\caption{An inner vertex for three geodesics}
\label{vertex}
\end{figure}

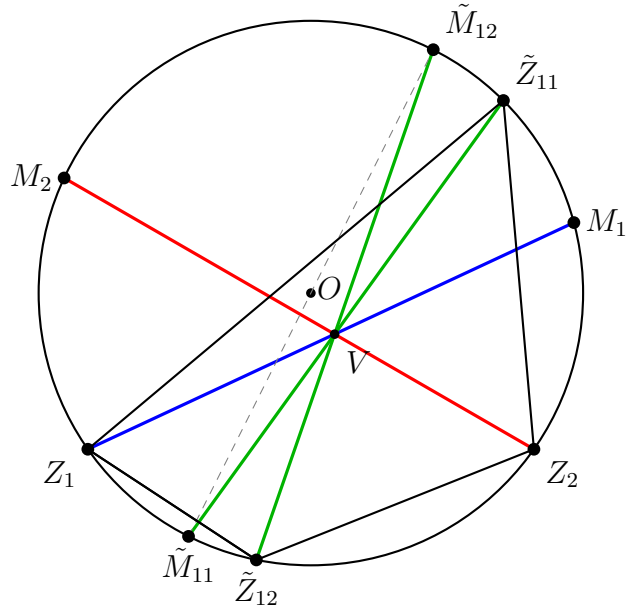
\begin{figure}[htbp]
\centering
\begin{tikzpicture}[scale=1.2, line cap=round, line join=round, thick]
    
    % Радиус окружности
    \def\R{3}
    
    % Окружность и центр O
    \draw[name path=circle] (0,0) circle (\R);
    \fill (0,0) circle (1.5pt) node[right, xshift=-2pt, yshift=2pt] {$O$};
    \coordinate (O) at (0,0);
    
    % Базовые точки на окружности
    \coordinate (Z1)  at (215:\R);
    \coordinate (M1)  at (15:\R);
    \coordinate (Z2)  at (325:\R);
    \coordinate (M2)  at (155:\R);
    \coordinate (Z11) at (45:\R);
    
    % Находим точку V как пересечение синей (Z1-M1) и красной (Z2-M2) линий
    \path[name path=L_blue] (Z1) -- (M1);
    \path[name path=L_red]  (Z2) -- (M2);
    \path[name intersections={of=L_blue and L_red, by=V}];

    % 1. Находим M11 (луч от Z11 через V)
    \path[overlay, name path=ray_Z11] ($(Z11)!-0.5!(V)$) -- ($(Z11)!3!(V)$);
    \path[overlay, name intersections={of=circle and ray_Z11, name=int_Z11}];
    \coordinate (M11) at (int_Z11-2);
    
    % 2. Находим M12 строго как диаметрально противоположную M11 через центр O
    \coordinate (M12) at ($(O)!-1!(M11)$);
    
    % 3. Находим Z12 (луч от нового M12 через V)
    \path[overlay, name path=ray_M12] ($(M12)!-0.5!(V)$) -- ($(M12)!3!(V)$);
    \path[overlay, name intersections={of=circle and ray_M12, name=int_M12}];
    \coordinate (Z12) at (int_M12-2);
    
    % === Отрисовка цветных линий (проходят через V) ===
    \draw[blue, very thick] (Z1) -- (M1);
    \draw[red, very thick]  (Z2) -- (M2);
    \draw[green!70!black, very thick] (Z11) -- (M11);
    \draw[green!70!black, very thick] (M12) -- (Z12);
    
    % Линия через центр О для наглядности (опционально, можно убрать)
    \draw[gray, dashed, thin] (M11) -- (M12);
    
    % === Отрисовка черных линий ===
    \draw (Z1) -- (Z12);
    \draw (Z1) -- (Z11);
    \draw (Z1) -- (Z12);
    
    \draw (Z11) -- (Z2); 
    \draw (Z12) -- (Z2);

    % === Явная отрисовка точек на окружности и подписи ===
    \fill (Z1)  circle (2pt) node[below left]  {$Z_1$};
    \fill (M1)  circle (2pt) node[right]       {$M_1$};
    \fill (Z2)  circle (2pt) node[below right] {$Z_2$};
    \fill (M2)  circle (2pt) node[left]        {$M_2$};
    \fill (Z11) circle (2pt) node[above right] {$\tilde{Z}_{11}$};
    \fill (M12) circle (2pt) node[above right]  {$\tilde{M}_{12}$};
    \fill (M11) circle (2pt) node[below] {$\tilde{M}_{11}$};
    \fill (Z12) circle (2pt) node[below]       {$\tilde{Z}_{12}$};
    \fill (V) circle (1.5pt) node[below right, yshift=-2pt] {$V$};
\end{tikzpicture}
\caption{Inner vertex. The $\mathbb C$-picture}
 \label{ditch}
\end{figure}

In this subsection we consider two geodesics, blue and red, that start on the boundary and reach a certain point in the bulk, where they merge and form an internal green geodesic. We artificially continue this geodesic up to the boundary to obtain two points $\tilde{Z}_{11}$ and $\tilde Z_{12} $\ref{vertex}. We map the crossing point to a point $V$ on a complex plane and obtain points $M_1, M_2, \tilde M_{11}, \tilde M_{12}$ \ref{ditch}.

$$\triangle Z_1 V\tilde{Z}_{12}\sim\triangle M_1 V\tilde{M}_{12}\Rightarrow \frac{Z_1\tilde{Z}_{12}}{M_1\tilde{M}_{12}}=\frac{V\tilde{Z}_{12}}{VM_1}$$
$$\triangle Z_2V\tilde{Z}_{12}\sim\triangle M_2 V\tilde{M}_{12}\Rightarrow \frac{Z_2\tilde{Z}_{12}}{M_2\tilde{M}_{12}}=\frac{V\tilde{Z}_{12}}{VM_2}$$
$$\triangle Z_1 V\tilde{Z}_{11}\sim\triangle M_1 V\tilde{M}_{11}\Rightarrow \frac{Z_1\tilde{Z}_{11}}{M_1\tilde{M}_{11}}=\frac{V\tilde{Z}_{11}}{VM_1}$$
$$\triangle Z_2V\tilde{Z}_{11}\sim\triangle M_2V\tilde{M}_{11}\Rightarrow \frac{Z_2\tilde{Z}_{11}}{M_2\tilde{M}_{11}}=\frac{V\tilde{Z}_{11}}{VM_2}$$
Multiplying the first and the last equality and dividing by the second and third ones, we get: 
\be  
\frac{Z_1\tilde{Z}_{12}\cdot Z_2\tilde{Z}_{11}}{Z_2\tilde{Z}_{12}\cdot Z_1\tilde{Z}_{11}}=\frac{M_1\tilde{M}_{12}\cdot M_2\tilde{M}_{11}}{M_2\tilde{M}_{12}\cdot M_1\tilde{M}_{11}}=\frac{\tan\frac{\angle M_1 O\tilde{M}_{12}}{2}}{\tan\frac{\angle M_2 O \tilde{M}_{12}}{2}}
\ee  
We can consider a triangle with sides $\epsilon_1, \epsilon_2, \tilde \epsilon_1$, since $\epsilon_1 e^{i\mu_1}+\epsilon_2 e^{i\mu_2}+\tilde \epsilon_1e^{i\tilde \mu_1}=0$. This means that in Fig.~\ref{halftan} $\angle M_1O\tilde M_{12}=\pi - \alpha_1, \;\; \angle M_2O\tilde M_{12}=\pi-\alpha_2$, and hence 
\be \label{cot}
\frac{Z_1\tilde{Z}_{12}\cdot Z_2\tilde{Z}_{11}}{Z_2\tilde{Z}_{12}\cdot Z_1\tilde{Z}_{11}}=\frac{\cot\frac{\alpha_1}{2}}{\cot\frac{\alpha_2}{2}}=\frac{-\epsilon_1+\epsilon_2+\tilde{\epsilon}_1}{\epsilon_1-\epsilon_2+\tilde{\epsilon}_1}
\ee
The last equality is depicted in Fig.~\ref{halftan}. 

The points $z_1, z_2, \tilde{z}_{11}, \tilde{z}_{12}$ lie on a circle, and therefore their cross-ratio must be real. Moreover, the segments $[Z_1 Z_2]$ and $[\tilde{Z}_{11}\tilde{Z}_{12}]$ have a nonempty intersection, so the cross-ratio must be negative. We thus obtain:
\be  \label{cross-ratio}
(\mathfrak w_1, \mathfrak w_2; \tilde{\mathfrak w}_{12}, \tilde{\mathfrak w}_{11})=\frac{(\mathfrak w_1-\tilde{\mathfrak w}_{12})(\mathfrak w_2-\tilde{\mathfrak w}_{11})}{(\mathfrak w_1-\tilde{\mathfrak w}_{11})(\mathfrak w_2-\tilde{\mathfrak w}_{12})}=-\frac{-\epsilon_1+\epsilon_2+\tilde{\epsilon}_1}{\epsilon_1-\epsilon_2+\tilde{\epsilon}_1}
\ee

\begin{figure}[htbp] 
\centering
\begin{tikzpicture}[scale=1, rotate=0, line cap=round, line join=round]

    % --- Точные математические координаты ---
    \coordinate (A) at (0,0);
    \coordinate (B) at (8,0);
    \coordinate (C) at (3.1875, 5.0833);
    
    % Координаты центра вписанной окружности (I) и точки касания (Tab)
    \coordinate (I) at (3.5, 1.9365);
    \coordinate (Tab) at (3.5, 0);

    % --- Отрисовка основных фигур ---
    % Треугольник
    \draw[thick] (A) -- (B) -- (C) -- cycle;
    
    % Вписанная окружность
    \draw[thick] (I) circle (1.9365);

    % Биссектрисы и радиус к нижней стороне
    \draw[thick] (A) -- (I);
    \draw[thick] (B) -- (I);
    \draw[thick] (I) -- (Tab);

    % --- Значок прямого угла ---
    % Квадратик 0.15 x 0.15 у точки касания
    \draw[thick] ($(Tab) + (0.15, 0)$) -- ($(Tab) + (0.15, 0.15)$) -- ($(Tab) + (0, 0.15)$);

    % --- Дуги углов (прописаны вручную для идеального отображения) ---
    % Левый угол (А): одна дуга (от 0° до 57.91°)
    \draw[thick] (A) ++(0:0.8) arc (0:57.91:0.8);
    
    % Правый угол (B): двойная дуга (от 180° до 133.43°)
    \draw[thick] (B) ++(180:0.8) arc (180:133.43:0.8);
    \draw[thick] (B) ++(180:0.9) arc (180:133.43:0.9);

    % --- Подписи углов (нормальный размер) ---
    \node at ($(A) + (14.5:1.15)$) {$\frac{\alpha_1}{2}$};
    \node at ($(A) + (43.5:1.15)$) {$\frac{\alpha_1}{2}$};
    
    \node at ($(B) + (168.4:1.25)$) {$\frac{\alpha_2}{2}$};
    \node at ($(B) + (145.2:1.25)$) {$\frac{\alpha_2}{2}$};

    % --- Подписи боковых сторон (нормальный размер) ---
    \node at ($(A)!0.5!(C) + (-0.35, 0.2)$) {$\epsilon_2$};
    \node at ($(B)!0.5!(C) + (0.35, 0.2)$) {$\epsilon_1$};

    % --- Подписи нижних отрезков ---
    % \displaystyle сохраняет нормальный вид горизонтальных дробей, но сам шрифт теперь базового размера
    \node[below=8pt] at ($(A)!0.5!(Tab)$) {$\frac{-\epsilon_1 + \epsilon_2 + \tilde{\epsilon}_1}{2}$};
    \node[below=8pt] at ($(Tab)!0.5!(B)$) {$\frac{\epsilon_1 - \epsilon_2 + \tilde{\epsilon}_1}{2}$};
    \node[above right] at (I) {$I$};

\end{tikzpicture}
\caption{Illustration of \ref{cot}}
\label{halftan}
\end{figure}
We are ready now to write down the action of, for example, the blue geodesic:
\be \label{action1}
S_1=\epsilon_1 \ln \frac{Z_1 V}{VM_1}=\epsilon_1\ln\frac{Z_1 V }{VM_2}\frac{VM_2}{V\tilde Z_{11}}\frac{V \tilde{Z}_{11}}{VM_1}=\epsilon_1 \ln\frac{Z_1Z_2}{M_1M_2}\frac{M_2\tilde M_{11}}{Z_2\tilde{Z}_{11}}\frac{Z_1\tilde Z_{11}}{M_1\tilde M_{11}}\sim \epsilon_1\ln\frac{(\mathfrak w_1-\mathfrak w_2)(\mathfrak w_1-\tilde {\mathfrak w}_{11})}{\mathfrak w_1(\mathfrak w_2-\tilde {\mathfrak w}_{11})}
\ee 
\subsection{5-point identity block}
The result for the 5-point vacuum block of type I ($\tilde \epsilon_1=\epsilon_3, \; \tilde\epsilon_2=0$) is obtained by summing the contributions of the three geodesics. Using the result \ref{action1}, we get:
\be \label{action geometry}
f(z)= -\sum\limits_{i=1}^3 \epsilon_i \ln z_i + \epsilon_1 \ln \frac{z_1^\alpha(z^\alpha_2-z^\alpha_3)}{(z^\alpha_1-z^\alpha_2)(z^\alpha_1-z^\alpha_3)}+ \epsilon_2 \ln \frac{z^\alpha_2(z^\alpha_1-z^\alpha_3)}{(z^\alpha_2-z^\alpha_1)(z^\alpha_2-z^\alpha_3)}+ \epsilon_3 \ln \frac{z^\alpha_3(z^\alpha_1-z^\alpha_2)}{(z^\alpha_3-z^\alpha_1)(z^\alpha_3-z^\alpha_2)}
\ee 
The results \ref{cross-ratio} and \ref{action geometry} literally reproduce \ref{cross} and \ref{action monodromy} from a geometric perspective, in the specific case of two equal heavy fields. These two building blocks will be useful for the general technique of computing heavy-light blocks. 
\section{Applications}
We now assemble these ingredients into a general algorithm and apply it to the computation of more complicated blocks. For simplicity, we perform the calculations for the case considered in Section 3, so that $\mathfrak w = z^{\alpha}$.
\subsection{4-point non-vacuum block}
A general 4-point non-vacuum block is obtained from a 5-point vacuum block by substituting the endpoints of the internal geodesic, which satisfy \ref{cross-ratio}:
\be \label{4-point cross-ration}
({\mathfrak w}_1, {\mathfrak w}_2; \tilde {\mathfrak w}_{12}, \tilde {\mathfrak w}_{11})=-\frac{-\epsilon_1+\epsilon_2+\tilde\epsilon_1}{\epsilon_1-\epsilon_2+\tilde\epsilon_1}
\ee
We also impose the condition that the angular momentum of the internal geodesic vanishes, and hence 
\be \label{p-condition}
\tilde {\mathfrak w}_{11}+\tilde {\mathfrak w}_{12}=0
\ee
Solving \ref{4-point cross-ration} and \ref{p-condition} together, we find
\be \label{Quadratic}
\tilde{z}_{11}^\alpha=
\tilde {\mathfrak w}_{11}=\frac{\beta ({\mathfrak w}_1-{\mathfrak w}_2)-\sqrt{\beta^2({\mathfrak w}_1-{\mathfrak w}_2)^2+4{\mathfrak w}_1{\mathfrak w}_2}}{2}, \;\; \beta =\frac{\epsilon_2-\epsilon_1}{\tilde \epsilon_1}
\ee 
The conformal block then reads:
\be 
\mathfrak F = z_1^{\Delta_1(\alpha-1)}z_2^{\Delta_2(\alpha-1)}\tilde{z}_{11}^{\tilde\Delta_1 \alpha }(z_1^\alpha-z_2^\alpha)^{-\Delta_2-\Delta_3+\tilde\Delta_1}(z_1^\alpha-\tilde z_{11}^\alpha)^{-\Delta_1+\Delta_2-\tilde \Delta_1}(z_2^\alpha-\tilde z_{11}^\alpha)^{\Delta_1-\Delta_2-\tilde{\Delta}_1}
\ee 
This is exactly the result obtained in \cite{Semiclassical, Holographic}.

%%%%%%%%%%%%%%%%%%%%%%%%%%%%%%%%%%%%%%%%%%%%%%%%%%%%%%%%%%%%%%%%%%%%%%%%%%%%%%%%%%%%%%%%%%%%%%%%%%%%%%%%%%%%%%%%%%%%%%%%%%%%%%%%%%%%%%%%%%%%%%%%%%%%%%%%%%%%%%%%%%%%%%%%%%%%%%%%%%%%%%%%%%%%%%%%%%%%%%%%%%%%%%%%%%%%%%%%%%%%%%%%%%%%%%%%%%%%%%%%%%%%%%%%%%%%%%%%%%%%%%%%%%%%%%%%%%%%%%%%%%%%%%%%%%%%%%%%%%%%%%%%%%
\subsection{5-point non-vacuum block}
In this subsection we consider the 5-point block and the corresponding geodesic network shown in Fig.~\ref{5point}. We continue the internal geodesics up to the boundary to obtain the points $\tilde {\mathfrak w}_{11}, \tilde {\mathfrak w}_{12}, \tilde {\mathfrak w}_{21}, \tilde {\mathfrak w}_{22}$. The four equations imposed on these points are obtained by applying \ref{cross} (equivalently, \ref{cross-ratio}) at each cubic vertex:
\begin{figure}[htbp]
\centering
\begin{tikzpicture}[line cap=round, line join=round, very thick]

    % Граница круга (AdS)
    \draw[ultra thick] (0,0) circle (4);

    % === Координаты внутренних вершин ===
    \coordinate (O) at (0,0);
    \coordinate (V1) at (0.4, -1.4);
    \coordinate (V2) at (1.5, 0);

    % === Координаты точек на границе ===
    \coordinate (Z1) at (225:4);
    \coordinate (Z12) at (255:4);
    \coordinate (Z2) at (315:4);
    \coordinate (Z22) at (0:4);
    \coordinate (Z3) at (12:4);
    \coordinate (Z11) at (30:4);
    \coordinate (Z21) at (180:4);

    % === Отрисовка линий (каналов) ===

    % 1. Темно-красная горизонтальная линия
    \draw[mydarkred, dashed] (Z21) -- (O);
    \draw[mydarkred] (O) -- node[above, text=black, yshift=2pt] {$\tilde{\epsilon}_2$} (V2);
    \draw[mydarkred, dashed] (V2) -- (Z22);

    % 2. Зеленая линия (\tilde{\epsilon}_1)
    % Пересчитанные углы для идеально плавной дуги без "крючков" на концах
    \draw[mygreen, dashed] (Z12) to[out=63, in=236] (V1);
    \draw[mygreen] (V1) to[out=56, in=229] node[left, text=black, xshift=-2pt, yshift=2pt] {$\tilde{\epsilon}_1$} (V2);
    \draw[mygreen, dashed] (V2) to[out=49, in=222] (Z11);

    % 3. Синяя дуга (\epsilon_1)
    \draw[myblue] (Z1) to[bend left=15] node[above left, text=black] {$\epsilon_1$} (V1);

    % 4. Красная дуга (\epsilon_2)
    \draw[myred] (V1) to[bend left=20] node[above right, text=black] {$\epsilon_2$} (Z2);

    % 5. Желтая дуга (\epsilon_3)
    \draw[myyellow] (V2) to[bend left=15] node[above right, text=black, yshift=2pt] {$\epsilon_3$} (Z3);

    % === Подписи точек ===
    \node[above left, xshift=2pt, yshift=-2pt] at (O) {\textsf{O}};
    \node[below left] at (Z1) {$Z_1$};
    \node[below] at (Z12) {$\tilde{Z}_{12}$};
    \node[below right] at (Z2) {$Z_2$};
    \node[right] at (Z22) {$\tilde{Z}_{22}$};
    \node[right] at (Z3) {$Z_3$};
    \node[above right] at (Z11) {$\tilde{Z}_{11}$};
    \node[left] at (Z21) {$\tilde{Z}_{21}$};

\end{tikzpicture}
\caption{5-point geodesic network}
 \label{5point}
\end{figure}
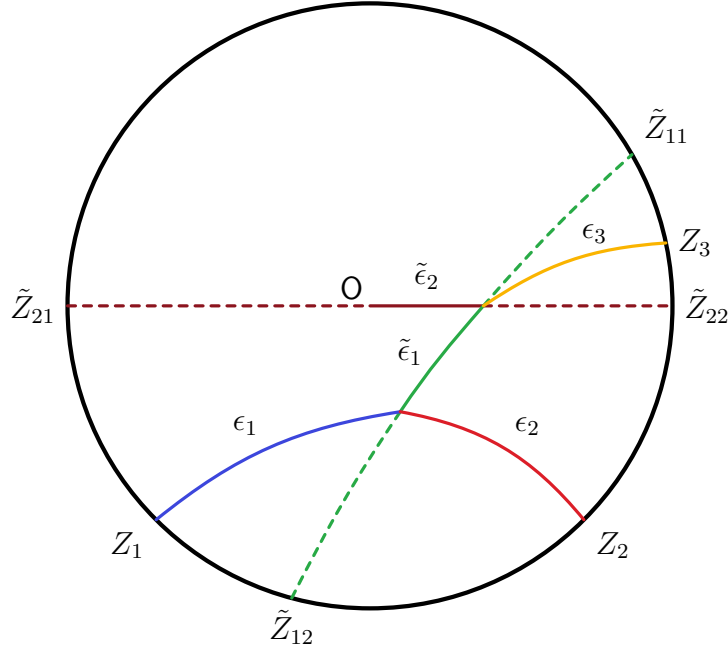

\be \label{equations}
\ba
(\mathfrak w_1, {\mathfrak w}_2; \tilde{{\mathfrak w}}_{12}, \tilde{{\mathfrak w}}_{11})=-\frac{-\epsilon_1+\epsilon_2+\tilde{\epsilon}_1}{\epsilon_1-\epsilon_2+\tilde{\epsilon}_1}, \;\;
({\mathfrak w}_3, \tilde{\mathfrak w}_{21}; \tilde{\mathfrak w}_{12}, \tilde{\mathfrak w}_{11})=-\frac{\tilde{\epsilon}_1+\epsilon_3-\tilde{\epsilon}_2}{\tilde{\epsilon}_1-\epsilon_3+\tilde{\epsilon}_2}
\\
(\tilde{\mathfrak w}_{12}, {\mathfrak w}_3; \tilde{\mathfrak w}_{22}, \tilde{\mathfrak w}_{21})=-\frac{-\tilde{\epsilon}_1+\epsilon_3+\tilde{\epsilon}_2}{\tilde{\epsilon}_1-
\epsilon_3+\tilde{\epsilon}_2}, \;\; (0, \infty; \tilde{\mathfrak w}_{22}, \tilde{\mathfrak w}_{21})=-1
\ea
\ee
where the last equation is somewhat formal, but it provides a convenient way of writing the condition. 

The action here consists of five contributions, from the blue, red, yellow, green, and brown geodesics. To compute the contribution of the green one, we use the following trick: the whole contribution can be split into three parts. The first is the whole geodesic without one dashed line, $\tilde{\epsilon}_1\ln \frac{({\mathfrak w}_1-\tilde{{\mathfrak w}}_{11})({\mathfrak w}_2-\tilde{{\mathfrak w}}_{11})}{\tilde{\mathfrak w}_{11} ({\mathfrak w}_1-{\mathfrak w}_2)}$; the second is the whole geodesic without the second dashed line, $\tilde{\epsilon}_1\ln \frac{({\mathfrak w}_3-\tilde{{\mathfrak w}}_{12})(\tilde{\mathfrak w}_{21}-\tilde{\mathfrak w}_{12})}{\tilde{\mathfrak w}_{12} ({\mathfrak w}_3-\tilde{\mathfrak w}_{21})}$; and we then have to subtract the length of the whole geodesic, $\tilde{\epsilon}_1\ln \frac{(\tilde{\mathfrak w}_{12}-\tilde{\mathfrak w}_{11})^2}{\tilde{\mathfrak w}_{11}\tilde{\mathfrak w}_{12}}$, because it has been counted twice. The final expression for the action reads:
\be 
\ba   
S = \epsilon_1\ln \frac{({\mathfrak w}_1-{\mathfrak w}_2)({\mathfrak w}_1-\tilde {\mathfrak w}_{11})}{{\mathfrak w}_1({\mathfrak w}_{2}-\tilde {\mathfrak w}_{11})}+
\epsilon_2\ln \frac{({\mathfrak w}_1-{\mathfrak w}_2)({\mathfrak w}_2-\tilde {\mathfrak w}_{11})}{{\mathfrak w}_2({\mathfrak w}_{1}-\tilde {\mathfrak w}_{11})}+\epsilon_3\ln \frac{({\mathfrak w}_3-\tilde {\mathfrak w}_{12})(\tilde {\mathfrak w}_{21}-{\mathfrak w}_{3})}{{\mathfrak w}_3(\tilde {\mathfrak w}_{21}-\tilde {\mathfrak w}_{12})}+
\\
+\tilde\epsilon_1\ln \frac{({\mathfrak w}_1-\tilde {\mathfrak w}_{11})({\mathfrak w}_2-\tilde {\mathfrak w}_{11}) (\tilde {\mathfrak w}_{21}-\tilde {\mathfrak w}_{12})({\mathfrak w}_3-\tilde {\mathfrak w}_{12})}{({\mathfrak w}_{1}-{\mathfrak w}_{2}) (\tilde {\mathfrak w}_{21}-{\mathfrak w}_3)(\tilde {\mathfrak w}_{11}-\tilde {\mathfrak w}_{12})^2}
+
\tilde\epsilon_2\ln \frac{(\tilde {\mathfrak w}_{21}-\tilde {\mathfrak w}_{12})(\tilde {\mathfrak w}_{21}-{\mathfrak w}_{3})}{\tilde {\mathfrak w}_{21}({\mathfrak w}_{3}-\tilde {\mathfrak w}_{12})} 
\ea 
\ee 
which can be rewritten in a more convenient invariant form:
\be \label{LongAction}
\ba 
S=\epsilon_1\ln \frac{(\mathfrak w_1-{\mathfrak w}_2)(\mathfrak w_1-{\mathfrak w}_3)}{\mathfrak w_1({\mathfrak w}_2-{\mathfrak w}_3)}+\epsilon_1&\ln (\mathfrak w_1,{\mathfrak w}_2; \tilde{\mathfrak w}_{11}, {\mathfrak w}_3)
+\epsilon_2\ln \frac{(\mathfrak w_1-{\mathfrak w}_2)({\mathfrak w}_2-{\mathfrak w}_3)}{{\mathfrak w}_2(\mathfrak w_1-{\mathfrak w}_3)}+\epsilon_2\ln ({\mathfrak w}_2, \mathfrak w_1; \tilde{\mathfrak w}_{11}, {\mathfrak w}_3)+\\
+\tilde{\epsilon}_1\ln ({\mathfrak w}_2, {\mathfrak w}_3; \tilde{\mathfrak w}_{11}, \tilde{\mathfrak w}_{12})+&\epsilon_3\ln \frac{(\mathfrak w_1-{\mathfrak w}_3)({\mathfrak w}_2-{\mathfrak w}_3)}{{\mathfrak w}_3(\mathfrak w_1-{\mathfrak w}_2)}+\epsilon_3\ln\Big(({\mathfrak w}_2, \tilde{\mathfrak w}_{12}; \tilde{\mathfrak w}_{21}, {\mathfrak w}_3)(\mathfrak w_1, \tilde{\mathfrak w}_{21}; {\mathfrak w}_2, {\mathfrak w}_3)\Big)
+\\+&\tilde{\epsilon}_2\ln\Big(({\mathfrak w}_3, 0; \tilde{\mathfrak w}_{21}, \tilde{\mathfrak w}_{12})(\tilde{\mathfrak w}_{21}, 0; \tilde{\mathfrak w}_{12}, \infty)\Big)
\ea 
\ee

To solve equations \ref{equations}, we make a global conformal map that sends $\mathfrak w_1\mapsto 0, {\mathfrak w}_2\mapsto 1, {\mathfrak w}_3\mapsto \infty$.
Then $\infty\mapsto x_\infty=\frac{{\mathfrak w}_2-{\mathfrak w}_3}{{\mathfrak w}_2-\mathfrak w_1}=\zeta+i\eta,\;\; 0 \mapsto x_0=\frac{\mathfrak w_1}{{\mathfrak w}_3}\frac{{\mathfrak w}_2-{\mathfrak w}_3}{{\mathfrak w}_2-\mathfrak w_1}=\zeta-i\eta$,
and we arrive at the following equation for $\tilde{x}_{11}=x$:
\be \label{polynomial}
P(x)=\Big(\zeta + (b -\zeta u) x + a x^2\Big)\Big(\zeta + (d - \zeta u) x + c x^2\Big)+\eta^2 (1 - u x)^2=0, 
\ee
where
\be \label{parameters}
\ba
u=\frac{2\tilde{\epsilon}_1}{\epsilon_1-\epsilon_2+\tilde{\epsilon}_1}, \; a=\frac{\tilde{\epsilon}_1-\epsilon_3+\tilde{\epsilon}_2}{\epsilon_1-\epsilon_2+\tilde{\epsilon}_1},\; b = \frac{-\epsilon_1+\epsilon_2+\epsilon_3-\tilde{\epsilon}_2}{\epsilon_1-\epsilon_2+\tilde{\epsilon}_1},\\
c = \frac{\tilde{\epsilon}_1-\epsilon_3-\tilde{\epsilon}_2}{\epsilon_1-\epsilon_2+\tilde{\epsilon}_1}, \; d = \frac{-\epsilon_1+\epsilon_2+\epsilon_3+\tilde{\epsilon}_2}{\epsilon_1-\epsilon_2+\tilde{\epsilon}_1}
\ea
\ee
We now need to decide which root corresponds to the physical solution. It is clear that the desired root is negative. We will prove that \ref{polynomial} has 4 real roots, with a single negative one. We carry out the proof for $\arg  \mathfrak w_i \in [0, \pi)$; in this case it is easy to see that $\zeta < 0$. We also denote
$$
Q_1=\Big(\zeta + (b -\zeta u) x + a x^2\Big), \;\; Q_2=\Big(\zeta + (d - \zeta u) x + c x^2\Big)
$$
Let us first take a closer look at $Q_1$: since $a>0$ and $\zeta < 0$, $Q_1$ has two real roots, one negative ($x_1$)  and one positive ($x_2$). Consider now $Q_2$: here $c<0$ and $\zeta < 0$, but $u >1$ and $c+d=u-1>0$, and hence $d>|c|$. Then 
$$
Q_2(u^{-1})=\frac{u d + c}{u^2}>\frac{u d - d}{u^2} > 0.
$$
It follows that $Q_2$ also has two real roots, with $0<x_3<u^{-1}$ and $u^{-1} < x_4$.  We now want to prove that $P(x)$ has a single negative root or, in other words, that the equation
$$
f(x)=\frac{Q_1(x)Q_2(x)}{(x-u^{-1})^2}=-u^2\eta^2
$$
has a single negative root. We first notice  that $\forall x \in (x_1, 0] \;\;f(x)>0$. Next we prove that on $(-\infty, x_1]$ there is a single root of $P(x)$. To do so, we consider
$$
\frac {f'}{f}=-\frac{2}{(x-u^{-1})}+\sum\limits_{i=1}^4 \frac{1}{(x-x_i)}<-\frac{2}{(x-u^{-1})}+\sum\limits_{i=1}^2 \frac{1}{(x-x_i)}<0
$$
because $x_{1,2}<0<u^{-1}$. In the interval $(-\infty, x_1]$ we have $f<0$, and therefore $f'>0$. Hence in this interval there is at most one root, which completes the proof. 

Knowing which root $x$ to pick, we can substitute it into the action and obtain: 
\be
\ba 
S=-\epsilon_1 \ln \mathfrak w_1-\epsilon_2 \ln \mathfrak w_2-(\epsilon_3+&\tilde\epsilon_2)\ln \mathfrak w_3+({\epsilon_1+\epsilon_2-\epsilon_3-\tilde{\epsilon}_2})   \ln(\mathfrak w_1-\mathfrak w_2)+(\epsilon_1-\epsilon_2+\epsilon_3+\tilde\epsilon_2)\ln(\mathfrak w_1-\mathfrak w_3)+\\&+
({-\epsilon_1+\epsilon_2+\epsilon_3+\tilde{\epsilon}_2})\ln(\mathfrak w_2-\mathfrak w_3)+G \big(\zeta, \zeta^2+\eta^2|\epsilon\big),
\\
G \big(\zeta, \zeta^2+\eta^2|\epsilon\big)=(\epsilon_1-\epsilon_2-\epsilon_3+&\tilde{\epsilon}_2)\ln x+({-\epsilon_1+\epsilon_2-\epsilon_3+\tilde{\epsilon}_2})\ln(1-x)+(\tilde{\epsilon}_1+\epsilon_3-\tilde{\epsilon}_2)\ln \Big(1-u x\Big)-\\
-\tilde{\epsilon_2}\ln \Big(&u x_\infty-a x+\frac{ub x}{1-u x}\Big)-\tilde\epsilon_2\ln \Big(u x_0-a x+\frac{u b x}{1-ux}\Big)
\ea
\ee 
And the conformal block reads:
\be 
\ba 
\mathfrak F \sim z_1^{(\alpha-1)\Delta_1}z_2^{(\alpha-1)\Delta_2}z_3^{(\alpha-1)\Delta_3+\alpha\tilde\Delta_2}(z_1^\alpha-z_2^\alpha)^{-\Delta_1-\Delta_2+\Delta_3+\tilde\Delta_2}(z_1^\alpha-z_3^\alpha)^{-\Delta_1+\Delta_2-\Delta_3-\tilde\Delta_2}\times\\ \times(z_2^\alpha-z_3^\alpha)^{\Delta_1-\Delta_2-\Delta_3-\tilde\Delta_2}\mathfrak G\big(\zeta, \zeta^2+\eta^2|\Delta, \tilde\Delta\big),\\
\mathfrak G\big(\zeta, \zeta^2+\eta^2|\Delta, \tilde\Delta\big)=x^{-\Delta_1+\Delta_2+\Delta_3-\tilde\Delta_2}(1-x)^{\Delta_1-\Delta_2+\Delta_3-\tilde{\Delta}_2}(1-ux)^{-\tilde\Delta_1-\Delta_3+\tilde\Delta_2}\times \\
 \times \Big(u x_\infty-a x+\frac{ub x}{1-u x}\Big)^{\tilde\Delta_2}\Big(u x_0-a x+\frac{u b x}{1-ux}\Big)^{\tilde\Delta_2}
\ea 
\ee 
A comment about the limits of this formula. One can consider the case $\epsilon_3=0, \; \tilde\epsilon_1=\tilde\epsilon_2$. Then \ref{polynomial} reduces to a quadratic equation, whose solution precisely corresponds to the solution of \ref{Quadratic}. The same is true for \ref{LongAction}: this action collapses to the action of the four-point block. 
%%%%%%%%%%%%%%%%%%%%%%%%%%%%%%%%%%%%%%%%%%%%%%%%%%%%%%%%%%%%%%%%%%%%%%%%%%%%%%%%%%%%%%%%%%%%%%%%%%%%%%%%%%%%%%%%%%%%%%%%%%%%%%%%%%%%%%%%%%%%%%%%%%%%%%%%%%%%%%%%%%%%%%%%%%%%%%%%%%%%%%%%%%%%%%%%%%%%%%%%%%%%%%%%%%%%%%%%%%%%%%%%%%%%%%%%%%%%%%%%%%%%%%%%%%%%%%%%%%%%
\subsection{General heavy-light block}

For a block with $n$ light fields we have $n-1$ internal geodesics and therefore $2n-2$ auxiliary endpoints if we continue them up to the boundary. Accordingly, we will have $2n-2$ equations on these points, which take the form 
\be \label{nequations}
\ba
(\mathfrak w_1, \mathfrak w_2; \tilde {\mathfrak w}_{12}, \tilde {\mathfrak w}_{11})&=-\frac{-\epsilon_1+\epsilon_2+\tilde{\epsilon}_1}{\epsilon_1-\epsilon_2+\tilde{\epsilon}_1}\;\;\; (A_2)\\
\\
(\tilde {\mathfrak w}_{k-2, 2}, {\mathfrak w}_k; \tilde {\mathfrak w}_{k-1, 2}, \tilde {\mathfrak w}_{k-1, 1})&=-\frac{-\tilde\epsilon_{k-2}+\epsilon_k+\tilde{\epsilon}_{k-1}}{\tilde\epsilon_{k-2}-\epsilon_k+\tilde{\epsilon}_{k-1}}, \;\; k = 3, \dots, n\;\;\; (A_k)\\
(\tilde {\mathfrak w}_{k-1, 1}, {\mathfrak w}_{k}; \tilde {\mathfrak w}_{k-2, 1},\tilde {\mathfrak w}_{k-2, 2})&=-\frac{-\tilde\epsilon_{k-1}+\epsilon_{k}+\tilde{\epsilon}_{k-2}}{\tilde\epsilon_{k-1}-\epsilon_{k}+\tilde{\epsilon}_{k-2}}, \;\; k=3, \dots, n\;\;\; (B_k)
\\
(\tilde {\mathfrak w}_{n-1, 1}, \tilde {\mathfrak w}_{n-1, 2}; 0, \infty)&= -1\;\;\; (B_{n+1})
\ea
\ee 
The last equality can be supported by the following argument. The block for one light field is given by $f_1(\mathfrak w_1|\epsilon_1)=\epsilon_1\ln\frac{\mathfrak w_1'}{\mathfrak w_1}$, and the corresponding accessory parameter reads $c_1=\frac{\psi_1'}{\psi_1}+\frac{\psi_2'}{\psi_2}$. From \ref{matrix} we get the following monodromy:
\be
M^{(1)}=
\epsilon_1 \begin{pmatrix}
 0& {\mathfrak w_1}^{-1}\\
 \mathfrak w_1& 0
\end{pmatrix}
\ee 
with eigenvectors $\begin{pmatrix}
    -1\\
    \pm \mathfrak w_1
\end{pmatrix}$.
Hence $\tilde{\mathfrak w}_{n-1,1}+\tilde{\mathfrak w}_{n-1, 2}=0$.

To reduce this system of equations to a single equation, we perform the following procedure. At the $k$-th step we express $\tilde {\mathfrak w}_{k-1,2}=\tilde {\mathfrak w}_{k-1,2}({\mathfrak w}_1, \dots, {\mathfrak w}_k, \tilde {\mathfrak w}_{11})$ from the equation $A_k$ and substitute it into the equation $B_{k+1}$. From $B_{k+1}$ we express $\tilde {\mathfrak w}_{k-1,1}=\tilde {\mathfrak w}_{k-1,1}({\mathfrak w}_1, \dots {\mathfrak w}_{k+1}, \tilde {\mathfrak w}_{11})$ and substitute it into the equation $A_{k+1}$. At the $n$-th step we obtain the desired single equation. At each step the degree of the equation doubles, so eventually we arrive at an equation for $\tilde {\mathfrak w}_{11}$ of degree $2^{n-1}$. 
Thus, when $n$ exceeds 3, the degree of the equation reaches 8 or higher, and the equation can no longer be solved analytically. On the other hand, the heavy background is explicitly known when the number of heavy fields is less than or equal to 3. Therefore, analytic solutions are available for blocks with at most 3 heavy and 3 light fields. 

Nevertheless, for the sake of completeness, we will complete the algorithm for computing the action. We define the vacuum 4-point light action $$f_2(\theta_1, \theta_2|\epsilon)=\Big(\epsilon\sum\limits_{i=1}^2\ln\mathfrak w_i-2\epsilon \ln(\mathfrak w_1-\mathfrak w_2)\Big)\Big|_{\mathfrak w_j=e^{i\theta_j}}$$ and the vacuum 5-point light action \be 
\ba 
f_3&(\theta_1,\theta_2,\theta_3|\epsilon_1,\epsilon_2,\epsilon_3)=\\
=&\bigg(\sum\limits_{i}^3\epsilon_i\ln \mathfrak w_i-(\epsilon_1+\epsilon_2-\epsilon_3)\ln(\mathfrak w_1-\mathfrak w_2)-(\epsilon_1-\epsilon_2+\epsilon_3)\ln(\mathfrak w_1-\mathfrak w_3)-(-\epsilon_1+\epsilon_2+\epsilon_3)\ln(\mathfrak w_2-\mathfrak w_3)\bigg)\bigg|_{\mathfrak w_j=e^{i\theta_j}}
\ea 
\ee
Then, for the whole geodesic network, one could naively compute the action as 
$$
f_3(\theta_2, \theta_1, \tilde \theta_{11}|\epsilon_2, \epsilon_1, \tilde\epsilon_1)+\sum\limits_{i=2}^{n-1}f_3(\theta_{i+1}, \tilde\theta_{i-1, 2}, \tilde\theta_{i, 1}|\epsilon_i, \tilde\epsilon_{i-2}, \tilde\epsilon_{i-1})
$$
In this sum, however, as is easy to see, the lines that connect the `imaginary' endpoints are counted twice. Subtracting them from the sum, we finally get:
$$
f(\theta|\epsilon)=f_3(\theta_1, \theta_2, \tilde \theta_{11}|\epsilon_1, \epsilon_2, \tilde\epsilon_1)+\sum\limits_{i=2}^{n-1} \Big(f_3(\theta_{i+1}, \tilde\theta_{i, 1},\tilde\theta_{i-1, 2}|\epsilon_{i+1}, \tilde\epsilon_{i}, \tilde\epsilon_{i-1})-f_2(\tilde\theta_{i-1,1},\tilde\theta_{i-1,2}|\tilde\epsilon_{i-1})\Big)
$$

%%%%%%%%%%%%%%%%%%%%%%%%%%%%%%%%%%%%%%%%%%%%%%%%%%%%%%%%%%%%%%%%%%%%%%%%%%%%%%%%%%%%%%%%%%%%%%%%%%%%%%%%%%%%%%%%%%%%%%%%%%%%%%%%%%%%%%%%%%%%%%%%%%%%%%%%%%%%%%%%%%%%%%%%%%%%%%%%%%%%%%%%%%%%%%%%%%%%%%%%%%%%%%%%%%%%%%%%%%%%%%%%%%%%%%%%%%%%

\section{Conclusion}
In this paper we connected the classical monodromy method to bulk geometry for
semiclassical Virasoro blocks in the heavy-light limit. Passing to holographic
coordinates $\mathfrak w=\frac{\psi_1^{(0)}}{\psi_2^{(0)}}$, we found
that the eigenvectors of the monodromy matrix encode the endpoints of the bulk
geodesics. This observation yields both the light action and the algebraic
equations governing the internal geodesic network, and, crucially, these equations
do not depend on the heavy background: the heavy and light sectors decouple. 

For two heavy operators we reproduced the same equations from elementary Euclidean geometry
on the plane, an independent derivation that makes the geometric content of the
monodromy data manifest. As an application we obtained the full non-vacuum
five-point HHLLL block, previously available only in the superlight approximation.

\begin{comment}
These ingredients combine into an algorithm. One first solves the algebraic system
that fixes the endpoints of the internal geodesic network, which we set up for
arbitrary $n$, and then reconstructs the classical action additively, as a sum of
elementary vacuum four- and five-point contributions --- a step we carry out explicitly
for the five-point block. The light network is determined by background-independent
equations, while the heavy operators enter only through an explicitly known
background. The same structure also determines the limits of the construction: analytic
solvability requires, in the light sector, that the network equations be solvable by
radicals and, in the heavy sector, that the background be available in closed form;
these two independent conditions single out the six-point HHHLLL block as the maximal
analytically tractable case. Beyond it, the method still applies and still defines
the block, but the answer is no longer expressible in elementary closed form.
\end{comment}

These ingredients combine into an algorithm. One first solves the algebraic system that fixes the endpoints of the internal geodesic network, which we set up for arbitrary $n$, and then reconstructs the classical action additively as a sum of elementary vacuum four-
and five-point contributions, which we carry out explicitly for the five-point block. The light network is determined by background-independent equations, while the heavy operators enter only through an explicitly known background.

The
same structure also determines the limits of the construction: analytic solvability requires, in the light sector, that the network equations be solvable by radicals and, in the heavy sector, that the background be available in closed form. Within our approach, these two independent conditions single out the HHHLLL six-point block as the largest configuration that admits a closed analytic treatment. Beyond it, the method still applies and defines the block, but the answer is no longer expressible in elementary closed form.

%%%%%%%%%%%%%%%%%%%%%%%%%%%%%%%%%%%%%%%%%%%%%%%%%%%%%%%%%%%%%%%%%%%%%%%%%%%%%%%%%%%%%%%%%%%%%%%%%%%%%%%%%%%%%%%%%%%%%%%%%%%%%%%%%%%%%%%%%%%%%%%%%%%%%%%%%%%%%%%%%%%%%%%%%%%%%%%%%%%%%%%%%%%%%%%%%%%%%%%%%%%%%%%%%%%%%%%%%%%%%%%%%%%%%%%%%%%%

\end{document}